\documentclass[copyright,creativecommons]{eptcs}
\providecommand{\event}{ThEdu 2023}

\usepackage{underscore}
\usepackage[T1]{fontenc}
\usepackage{subcaption}
\usepackage{graphicx}
\usepackage{amsmath}
\usepackage{amssymb}
\usepackage{cite}
\usepackage{cleveref}

\RequirePackage{array}
\newenvironment{authors}[1]%
  {\begingroup
   \newcommand\estyle{}%
   \renewcommand\institute[1]%
     {\\\multicolumn{#1}{@{}c@{}}{\scriptsize\begin{tabular}[t]{@{}>{\footnotesize}c@{}}##1\end{tabular}}}%
   \renewcommand\email[1]%
     {\gdef\estyle{\footnotesize\ttfamily}\\##1\gdef\estyle{}}
   \begin{tabular}[t]{@{}*{#1}{>{\estyle}c}@{}}
  }%
  {\end{tabular}%
   \endgroup
  }

\title{Teaching Higher-Order Logic Using Isabelle}

\author{
  \begin{authors}{2}
    Simon Tobias Lund & Jørgen Villadsen
      \institute{Technical University of Denmark, Kongens Lyngby, Denmark}
      \email{sitlu@dtu.dk & jovi@dtu.dk}
  \end{authors}
}

\def\authorrunning{S. T. Lund \& J. Villadsen}
\def\titlerunning{Teaching Higher-Order Logic Using Isabelle}

\begin{document}
\maketitle

\begin{abstract}
We present a formalization of higher-order logic in the Isabelle proof assistant, building directly on the foundational framework Isabelle/Pure and developed to be as small and readable as possible. It should therefore serve as a good introduction for someone looking into learning about higher-order logic and proof assistants, without having to study the much more complex Isabelle/HOL with heavier automation. To showcase our development and approach we explain a sample proof, describe the axioms and rules of our higher-order logic, and discuss our experience with teaching the subject in a classroom setting.
\end{abstract}

\section{Introduction}

Higher-order logic, also known as simple type theory \cite{FARMER2008267}, has been described as the combination of functional programming and logic \cite{prog-prove}, and has proved a very powerful tool for the formalization of mathematics and computer science.
It is an expressive enough logic to cover a wide array of fields, while still being built on relatively simple principles, and a number of proof assistants based on higher-order logic are available.

We consider formal reasoning in the generic proof assistant Isabelle  \cite{Isabelle02,paulson}.
In the present paper we are taking advantage of the genericity of Isabelle, but we also find that Isabelle is at least as user-friendly and intuitive as other proof assistants of comparable power.
Although Isabelle is generic and comes with a number of object logics like first-order logic (FOL) and axiomatic set theory (ZF), the default object logic is higher-order logic, called Isabelle/HOL.
One of the main aims of the present work is to gently introduce students to Isabelle/HOL and its tutorials \cite{Isabelle02,prog-prove}. 

In \cite{ThEdu21}, we gave an Isabelle formalization of intuitionistic and classical propositional logic as a natural deduction system. That formalization is here expanded to higher-order logic, and thereby also first-order logic. In contrast to some other Isabelle formalizations of proof systems \cite{NaDeA18,LSFA}, which are developed in the complex Isabelle/HOL, we give a theory built directly on the fundamental Isabelle/Pure theory. Our formalization can therefore be viewed as an alternative (though very minimal) logical foundation to Isabelle/HOL. The main purpose of the formalization is as a teaching tool. It is small and transparent enough for someone with a working understanding of logic, but no experience with Isabelle, to study and understand thoroughly, something which is in practice very difficult with Isabelle/HOL.

The rest of the paper is structured as follows. We introduce our approach and describe our contributions in Section 2. We discuss related work in Section 3.
In Section 4, we go through sample natural deduction proofs in detail, for propositional logic, first-order logic and higher-order logic, as an introduction to Isabelle code. We present formalizations of first-order logic in Section 5 as a stepping stone towards higher-order logic. In Section 6 and Section 7, we build up intuitionistic and classical higher-order logic, respectively. Finally, in Section 8, we discuss further developments and provide concluding remarks.

\section{Approach and Contributions}

Higher-order logic is often considered fundamentally different from first-order logic.
This point of view is called a dogma in a recent paper \cite{10.1145/3557998}:
\begin{quote}
\emph{But now there is a new generation of higher-order provers that aim to gracefully generalize their first-order counterparts. They give up the dogma that higher-order logic is fundamentally different from first-order logic. Instead, they regard first-order logic as a fragment of higher-order logic for which we have efficient methods, and they start from that position of strength.}
\end{quote}
We find that teaching higher-order logic should be a graceful extension of teaching first-order logic, like the approach for provers \cite{10.1145/3557998}:
\begin{quote}
\emph{Accordingly, a strong higher-order prover should behave like a first-order prover on first-order problems, perform mostly like a first-order prover on mildly higher-order problems, and scale up to arbitrary higher-order problems.}
\end{quote}
In the present paper we often restrict ourselves to the first-order features with respect to axioms, rules and theorems, but always with the power of the higher-order features when needed.

Our main contributions are as follows:
\begin{itemize}
\item Our approach provides a gentle introduction to Isabelle/HOL, with many simplifications and no automation, but acknowledging the importance of higher-order logic in automated reasoning in general and in proof assistants in particular, for mathematics and computer science.
\item Our approach provides a natural deduction alternative to the foundation of mathematics, which is often based on axiomatic set theory, and the higher-order logic formalizes the axioms of choice, infinity and set comprehension.
\item Our approach includes a formalization of implicational axiomatics, with classical implicational logic as a propositional logic fragment, as a common starting point for our teaching, with formal proofs of the soundness and completeness theorems in Isabelle/HOL.
\item Our approach includes a succinct formalization of implication and universal quantification using the tools of Isabelle/Pure.
\item Our approach has been used, successfully, four times in a computer science course on automated reasoning in the period 2020-2023 for about one hundred students in total.
\end{itemize}
There are three different ``versions'' of the Isabelle system we will consider in this paper:
\begin{enumerate}
    \item Isabelle/Pure
    \item Isabelle/HOL, and
    \item Isabelle/HOL\_Pure.
\end{enumerate}
Isabelle/Pure contains the fundamentals of the Isabelle framework, like proof states, the proof environment, and the meta-logical operators like $\Longrightarrow$ (meta-logical implication) and $\bigwedge$ (meta-logical universal quantification). Isabelle/HOL is the main Isabelle system which contains a wide range of theories for many different fields of mathematics and computer science. Isabelle/HOL\_Pure is our minimal formalization of higher-order logic. It contains exactly the same axioms as Isabelle/HOL, but only a minimum of extra definitions and derived rules. Both Isabelle/HOL and Isabelle/HOL\_Pure import Isabelle/Pure. 

We consider first-order logic systems FOL\_Natural\_Deduction and FOL\_Implicational\_Axiomatics, where the first is based on natural deduction while the other is axiomatic. Both systems are formalized as deep embeddings (that is, we define the formulas as an object-level datatype) in Isabelle/HOL, where we have proved them sound and complete. The simple higher order logic is HOL\_Pure where we also have some example proofs and derived rules. Because we are axiomatizing this logic right on top of the Isabelle framework we cannot prove meta-properties like soundness or completeness of this system, like we could with the deep embeddings. If one tried to show soundness and completeness of the higher-order logic one would have to be able to define and reason about the semantics of the formulas, which is not possible because the structure of formulas is not accessible at the object level. The idea is to use both of these approaches to teach the students about Isabelle/HOL. The first-order logic systems are broad examples of somewhat complicated formalization efforts built on the full Isabelle system. They showcase patterns and techniques the students can later use in their own formalizations. The goal of the minimal HOL\_Pure system is somewhat different. Here we aim to give the students a very thorough understanding of the most fundamental concepts in Isabelle, as opposed to a shallow understanding of a much larger portion. 

These three formalizations are contained in the following Isabelle files:
\begin{itemize}
    \item[]\textbf{FOL\_Implicational\_Axiomatics.thy} A sound and complete deep embedding of an axiomatic proof system for first-order logic. The file contains 829 lines and is based on the entries Implicational\_Logic and FOL\_Axiomatic in Isabelle's Archive of Formal Proofs.
    \item[]\textbf{FOL\_Natural\_Deduction.thy} A sound and complete deep embedding of a natural deduction  proof system for first-order logic. The file contains 1386 lines and is based on the entry Synthetic\_Completeness in Isabelle's Archive of Formal Proofs.
    \item[]\textbf{HOL\_Pure.thy} A minimal formalization of higher-order logic built directly on the Isabelle/Pure framework. The file contains 666 lines and is based on the Isabelle/Pure examples by Makarius Wenzel.
\end{itemize}
Our formalizations are available online:
\begin{center}
    \url{https://hol.compute.dtu.dk/Pure/}
\end{center}
We have used our approach in our automated reasoning course since 2020:
\begin{center}
    \url{https://kurser.dtu.dk/course/02256}
\end{center}
This is an advanced MSc course (about 40 students) with a focus on the natural deduction proof system, first-order logic, higher-order logic and type theory, in particular Isabelle/HOL~\cite{Isabelle02}.

The course is a 5 ECTS point MSc course and we have described various aspects of the course elsewhere \cite{FMTea,ThEdu21,ThEdu20,EPTCS375.6,SLAI} but we are not aware of other related work for teaching higher-order logic and natural deduction.
In the course the students hand in 5 assignments and finally there is a 2-hour exam where the part using the approach described here usually constitutes 30\%\ of the exam.
All problems must be solved using the Isabelle proof assistant.
The Isabelle files are handed in online.

In the course we are teaching we have some overarching goals. The students should gain a broad understanding of logic and reasoning, with formal proofs in higher-order logic as the end goal. Tied in with this is interactive theorem proving and foundations of mathematics. They should both learn how to use the Isabelle proof assistant in practice and the broader context of logic in computer science and the general project of formalizing mathematics. 

For each of our aims we try to build up understanding in stages, starting with concepts that are easier or more familiar to the students. For example, we start by teaching propositional logic, then build up to first-order logic, and finally higher-order logic. In this paper we examine one such pipeline going from formalized first-order logic systems, to a formalized simple higher-order logic, and then finally the full Isabelle/HOL.

\section{Related Work}

In this paper, our main goal is to show how fundamental concepts of Isabelle can be taught using a simplified axiomatization of the same higher-order logic used in Isabelle/HOL.

In \cite{bohnelearning2018}, a description of using Coq for teaching formal proofs to students is given. They make a connection to ``textbook'' proofs and how learning Coq first might make the students better at correct but more informal proofs later on. They give some example proofs similar in scope to example proofs and derived rules given in this paper. 

In \cite{korkutproof2023}, a software tool for helping students construct proof trees in an interactive way is described. \cite{vasconcelosanita2023} describes a tool for verifying student proofs in tableaux. We view Isabelle as an alternative to such proof guides and checkers as it contains facilities both for guiding (and in some cases fully automating) the proof search and for checking the correctness of proofs. 

In \cite{villadsennatural2018}, a first-order logic natural deduction proof system is given. It is defined as a deep embedding in Isabelle/HOL and shown sound and complete. That paper focuses on using this system for teaching first-order logic to students. In addition to their Isabelle implementation they also give an interactive website for building and exporting proofs. The higher-order natural deduction developed in this paper is of course much more expressive but comes at the cost of verified meta properties like soundness and completeness. 

We focus on proof systems based on implication, but it is worth noting that classical propositional logic can be presented, say, in a language containing only disjunction and negation as primitive connectives.
For example, we have elsewhere \cite{CICM21} formalized the following axiom system with modus ponens as the only rule:
\begin{enumerate}
\item 
        \( p \vee p \rightarrow p \)
\item         
        \( p \rightarrow p \vee q \)
\item         
        \( (p \rightarrow q) \rightarrow (r \vee p) \rightarrow (q \vee r) \)

\end{enumerate}
Here $p \rightarrow q$ is an abbreviation for $\neg p \vee q$.
Without this abbreviation the axioms and the modus ponens rule would be difficult to grasp.
The axiom system was first proved sound and complete by Rasiowa in 1947 \cite{CICM21}, building on work by Russell and others.
For natural deduction, in particular as in Isabelle/HOL, we find that the focus on implication is best, given the meta-logical operators like $\Longrightarrow$ (meta-logical implication) mentioned in the previous section.

\section{Sample Natural Deduction Proofs in Isabelle}

Our approach to teaching higher-order logic involves natural deduction proofs in propositional logic (see Subsection 4.1) and first-order logic (see Subsection 4.2), because a solid basis is needed. Finally, in Subsection 4.3, we consider natural deduction proofs of Cantor's theorem, a well-known result going beyond first-order logic.

\subsection{Propositional Logic}

We give a brief introduction to proving in Isabelle by the following sample natural deduction proof, which is an improved version of a proof given in our previous paper \cite{ThEdu21}.
The proof works as-is in both our Isabelle/HOL\_Pure and in standard Isabelle/HOL.

\begin{center}
\includegraphics[width=\textwidth]{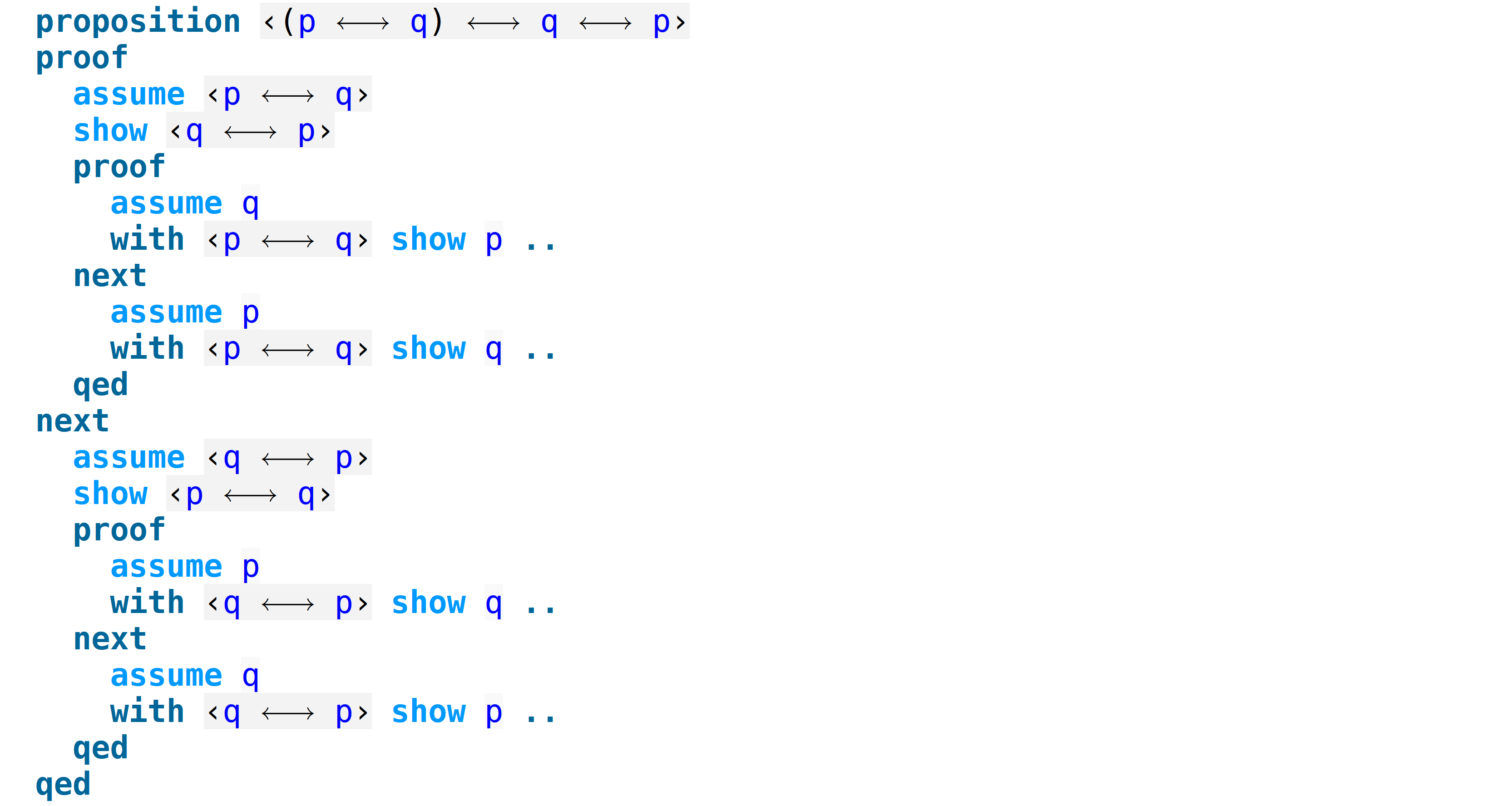}
\end{center}

The above figure shows a proof of the formula $(p \longleftrightarrow q) \longleftrightarrow (q \longleftrightarrow p)$.
Note that the second pair of parentheses in the formula $(p \longleftrightarrow q) \longleftrightarrow (q \longleftrightarrow p)$ is dropped in the Isabelle proof, as all arrows are right-associative. 

The structure of the proof should look familiar for logicians who know natural deduction. There are just two layers to the proof and four terminal branches. 
In the first layer we show the main bi-implication by assuming the left bi-implication then deriving the right bi-implication, then vice versa. The proofs of these two bi-implications make up the second layer. These proofs consist of assuming $p$ then showing $q$, and vice versa. 

The keywords of Isabelle and the Isar proof language \cite{Wenzel99,Wenzel2007,isar-ref} are similar to what one would use in a pen-and-paper natural deduction proof. 

The starting ``proposition'' introduces a terminal fact which, once proved, becomes available for the remaining theory (and any other Isabelle documents that import this theory). There are several synonyms to ``proposition'', like ``lemma'', ``theorem'', and ``corollary''. After the ``proposition'' comes the formula which we want to prove. 

We then use the ``proof'' keyword, which applies a single rule (in this case Isabelle has chosen the bi-implication introduction rule) and starts an Isar environment. The alternative ``proof -'' does the same without applying any rule first. The bi-implication introduction rule transforms the proof-state from the given formula to the two formulas: $p \longleftrightarrow q \Longrightarrow q \longleftrightarrow p$ and $q \longleftrightarrow p \Longrightarrow p \longleftrightarrow q$ (though the $\longleftrightarrow$ is displayed as $=$ in the Isabelle proof-state). $\Longrightarrow$ corresponds to a meta-implication in Isabelle, as opposed to the object-implication $\longrightarrow$. The meta-implications inform what the proof in the Isar environment should look like: the formula to the right of the rightmost arrow must be shown, while all the other formulas can be assumed. 

The Isar environment consists of a list of facts, corresponding to proofs of these two formulas. In general, these facts will either be introduced by ``assume'', ``have'', or ``show'', though only ``assume'' and ``show'' are used in this proof. A fact introduced by ``assume'' requires no proof, while facts introduced by ``have'' and ``show'' must be shown either directly (using a possibly empty list of rules and proof methods) or by introducing new proof-blocks. This technique of introducing new proof-blocks for facts stated in another proof-block leads to the nested nature of Isabelle proofs. A fact introduced by ``show'' must finish off a goal of the proof-state, while a fact introduced by ``have'' can be any desired intermediate property. In each layer of the given proof we simply assume one formula and show another. 

Since all the proofs introduced by the bi-implication introduction rule require showing two formulas, we need to clear the state after showing the first (thereby forgetting the previous assumptions) by using the ``next'' keyword. 

There are four terminal facts in our proof: $p$, $q$, $q$, and $p$. We tell Isabelle to prove these directly by using ``..'', which looks for an appropriate rule. In this case we need additional facts to match the bi-implication elimination rule. These facts are imported by the ``with'' keyword, which adds both explicitly given formulas and the previous fact to the context.

\subsection{First-Order Logic}

In our formalizations we have included 25 exercises covering mainly propositional and first-order logic.
The exercises are of varying difficulty.

We include the solution to the final exercise below as a showcase for the power and elegance of our HOL\_Pure formalization.

\begin{center}
\includegraphics[width=\textwidth]{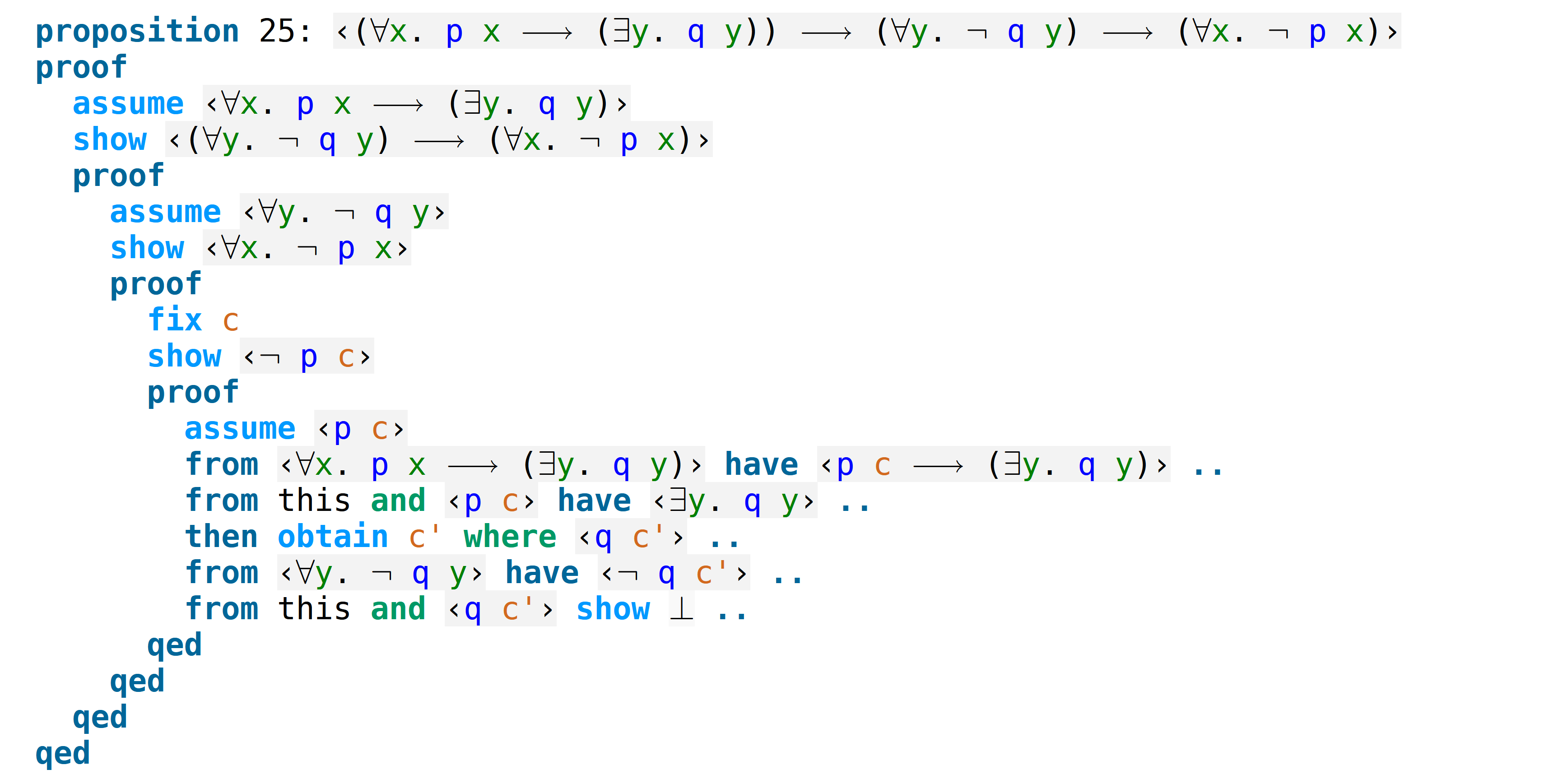}
\end{center}

The first two ``proof'' statements apply the implication introduction rule to break down the object-level implication to the meta implication, thereby allowing us to ``assume'' the precedent and ```show'' the somewhat simpler antecedent. 

The third ``proof'' statement applies the universal introduction rule, so that we only have to show the statement for some specific but arbitrary $c$. The last proof statement applies the negation introduction rule, which allows us to assume the non-negated statement and try to derive a contradiction.

In the final proof block we use universal elimination, modus ponens, and existential elimination over the facts we have assumed so far, to obtain a $c'$ for which $q$ both does and does not hold. This finishes the proof.

\subsection{Higher-Order Logic}

Cantor's theorem is a famous result in set theory:
\begin{quote}
\medskip

There is no surjective function from a set to its power set.    

\medskip
\end{quote}
Several formulations of Cantor's theorem are possible in higher-order logic. We have chosen to consider the following formulation in Isabelle/HOL:

\begin{center}
\includegraphics[width=\textwidth]{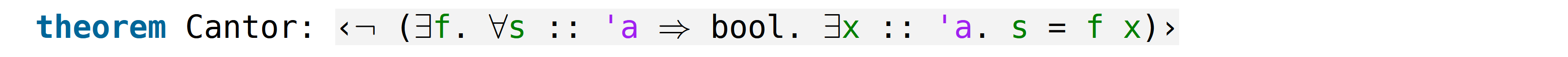}
\end{center}

The notation $'a$ is a stand in for an arbitrary type (non-empty).
Note that sets can be represented as functions from the type of the elements to $bool$. We say that an element is in such a ``set'' if the function returns true for it.

It is illustrative for students to see that the proof of Cantor's theorem can be carried out in intuitionistic higher-order logic.
Here is the full proof:

\begin{center}
\includegraphics[width=.8\textwidth]{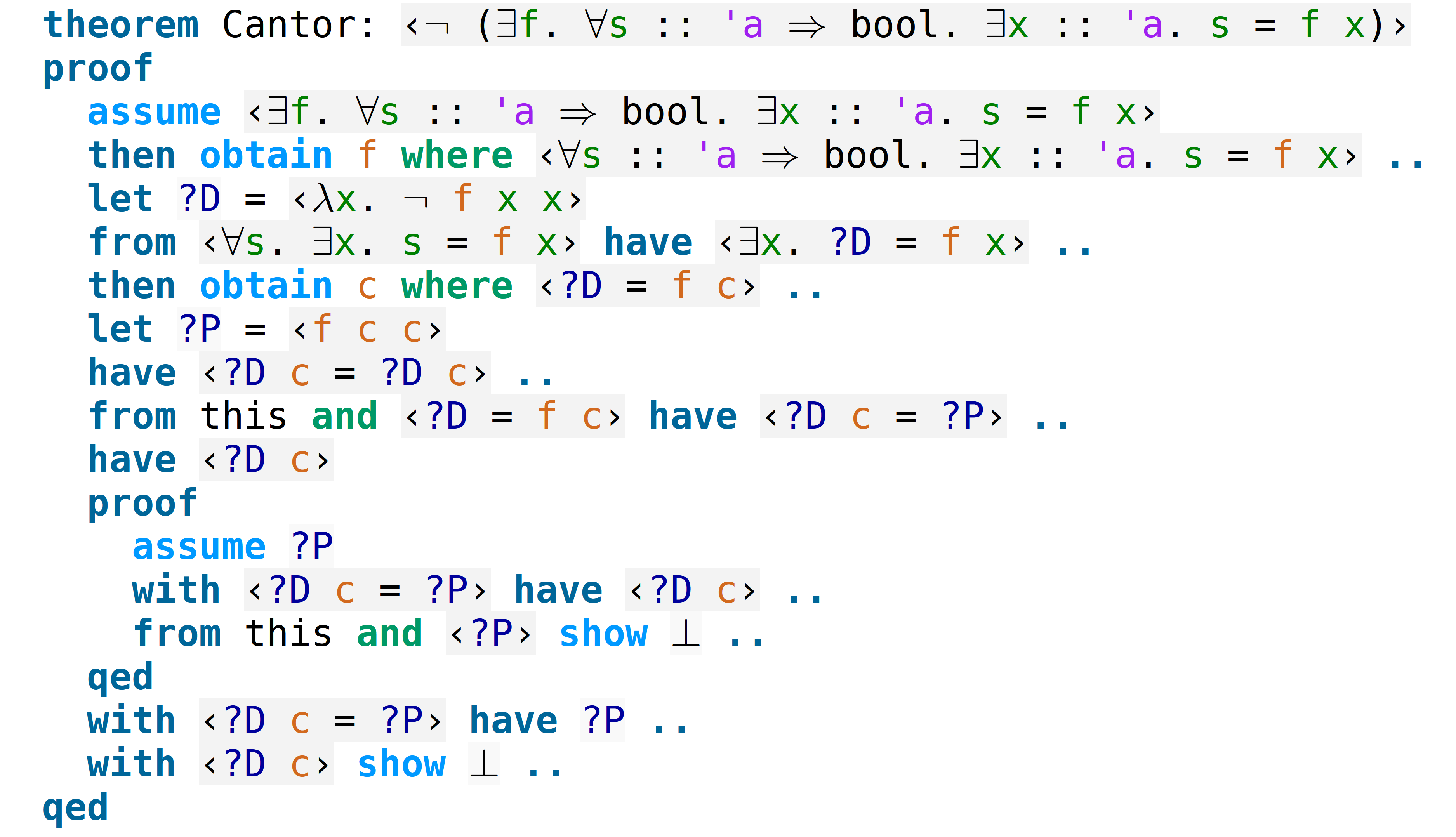}
\end{center}

The proof above is written completely without appeal to advanced proof tactics, and each line represents at most one rule application. It thus works in both Isabelle/HOL\_Pure and Isabelle/HOL (though in the latter one has to add $\bot$ as alternative notation for $False$). A detailed description of the proof follows. 

\begin{center}
\includegraphics[width=.8\textwidth]{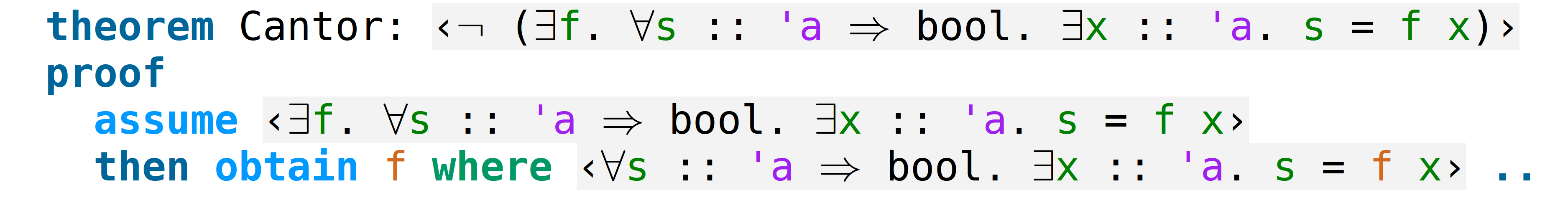}
\end{center}

We use negation introduction (automatically picked by ``proof'') to transform the proof state from $\neg (\exists f.\ \forall s.\ \exists x.\ s = f\ x)$ to $\exists f.\ \forall s.\ \exists x.\ s = f\ x \Longrightarrow \bot$. Note that negation introduction actually just unfolds the definition of negation, and is therefore still intuitionistic. In the new proof state, we can assume the existence of a surjective function from the type $'a$ to $'a \Rightarrow bool$ (i.e. from a set to its power set). The ``obtain'' statement allows us to reference this function in the rest of the proof. The proof state now requires us to show $\bot$, i.e. find a contradiction. 

\begin{center}
\includegraphics[width=.8\textwidth]{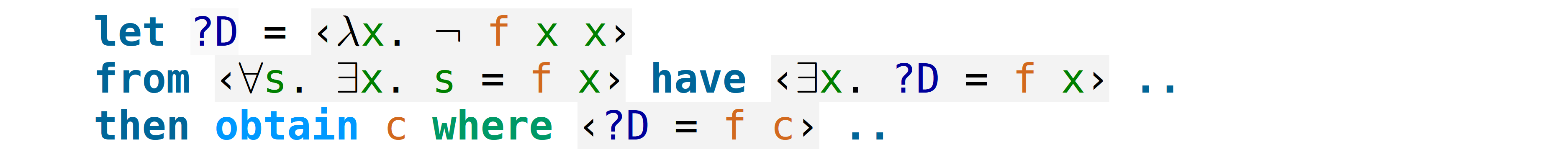}
\end{center}

For some $x$ of type $'a$, $f\ x\,$ will be a set where the elements also have type $'a$. We define a new set, $?D$, containing exactly those $x$ not contained in $f\ x\,$ (recall that $f\ x\ x\,$ should be interpreted as the statement ``$x$ is in $f\ x$'').

We can show by our definition of $f$ that there must be some $c$ such that $?D = f\ c$. 

\begin{center}
\includegraphics[width=.8\textwidth]{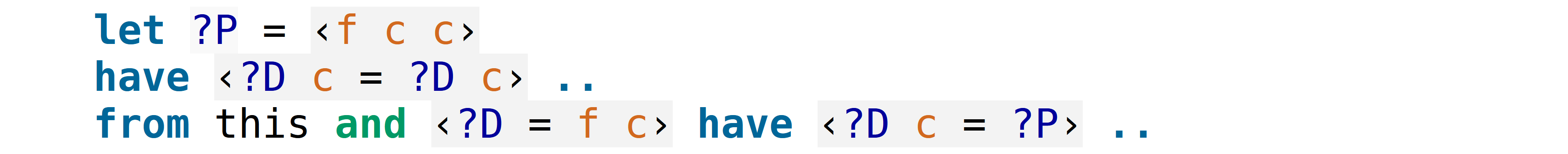}
\end{center}

We let $?P$ be the statement that $c$ is in $f\ c$. By reflexivity of equality and substitution we can prove that $?P$ must equal $c$ being in $D$. 

\begin{center}
\includegraphics[width=.8\textwidth]{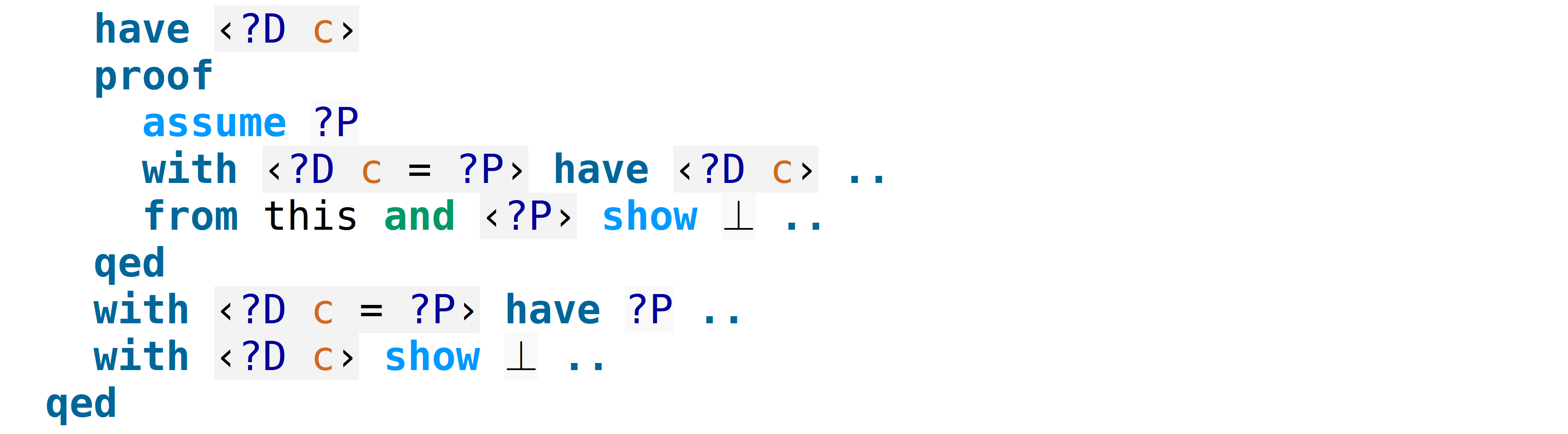}
\end{center}

This leads to a contradiction we can use to finish the proof. First, we can show that $c$ must be in $?D$. Since this corresponds to $c$ not being in $f\ c$, we can start by assuming $?P$ and try to show $\bot$. By the previous proof step we know that this must mean that $c$ is in $?D\ c$. When unfolding the abbreviations we get that $c$ both is and is not in $f\ c$, which shows $\bot$.

This shows that $c$ must be in $?D$. By the same line of reasoning as above (though using the equality between $c$ being in $?D$ and $?P$ holding in the opposite direction) we get that $c$ both is and is not in $f\ c$. This finishes the proof. 

We find that the full proof above is very useful for the students due to the lack of automation, but students should also work with the automation available in Isabelle/HOL, like in the following alternative proof.

\begin{center}
\includegraphics[width=.8\textwidth]{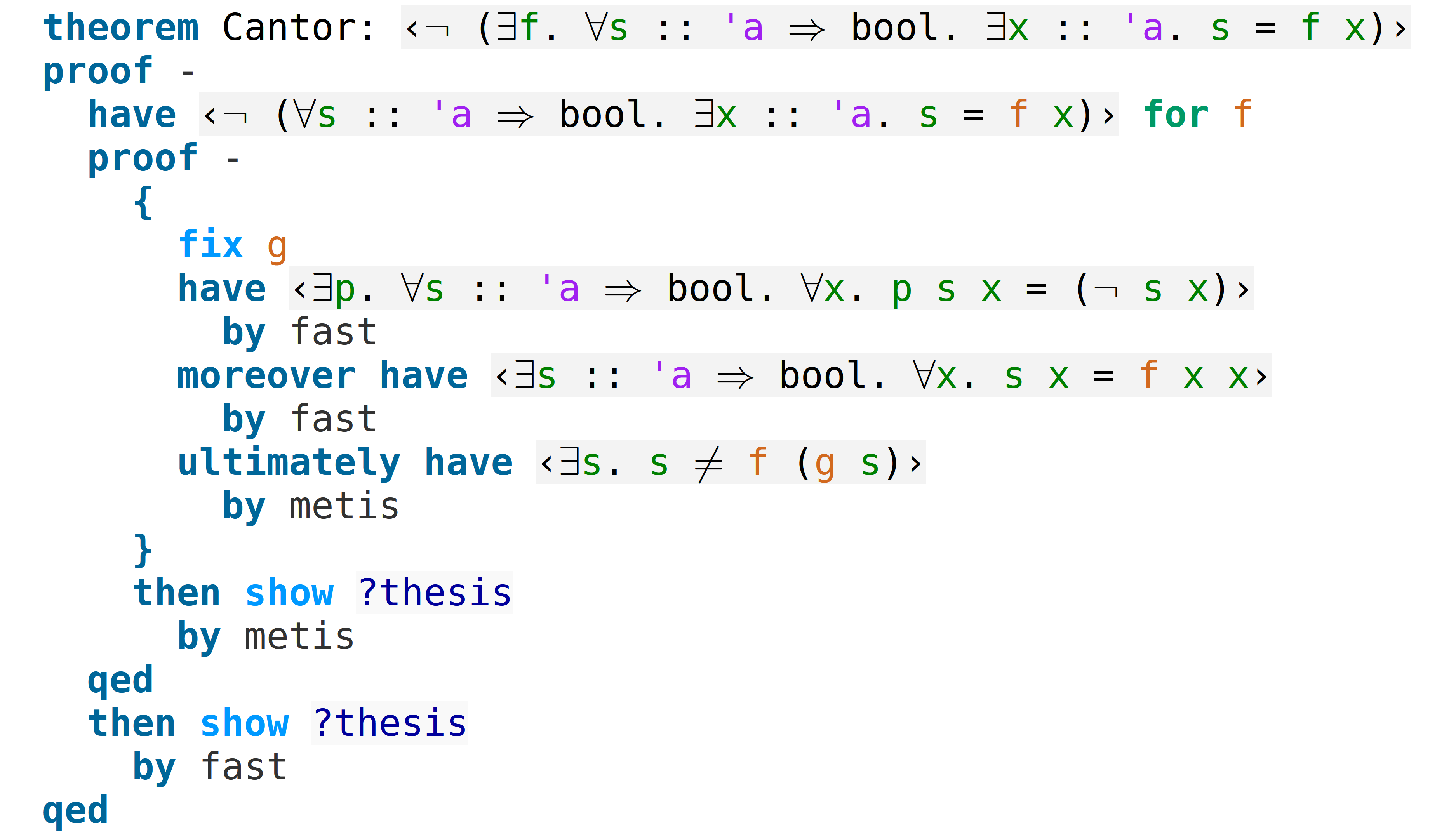}
\end{center}

The alternative proof above was found automatically by sledgehammer. We have rewritten the output of sledgehammer somewhat to be more legible, but not more than what we expect any student with a rudimentary understanding of Isabelle would be able to. Sledgehammer was only able to find the proof when the theorem is stated with $f$ as a schematic variable, instead of a bound variable as in the original formulation. This is why we first prove an alternative formulation of the theorem, and then show that the quantified version follows from this. 

We note that Cantor's theorem is right on the edge of what Isabelle's automation is able to handle. Sledgehammer could not return a single tactic application to solve the proof state, as is the common usage of this tool, but rather gave a large and quite illegible Isar proof. Furthermore, we were only able to find this proof using sledgehammer when using an unstable 2024 snapshot of Isabelle instead of the stable 2023 release. 

In the stable 2023 release we obtain the following message from sledgehammer:

\begin{verbatim}
        zipperposition found a proof... 
        zipperposition: One-line proof reconstruction failed: by metis
        Warning: Isar proof construction failed 
        Done
\end{verbatim}

In the snapshot available 10 January 2024 the Isar proof construction no longer fails.
However, instead we obtained the illegible Isar proof mentioned above.

By a thorough examination of the proof given by sledgehammer, one can reduce it to the following two-step proof. 

\begin{center}
\includegraphics[width=.8\textwidth]{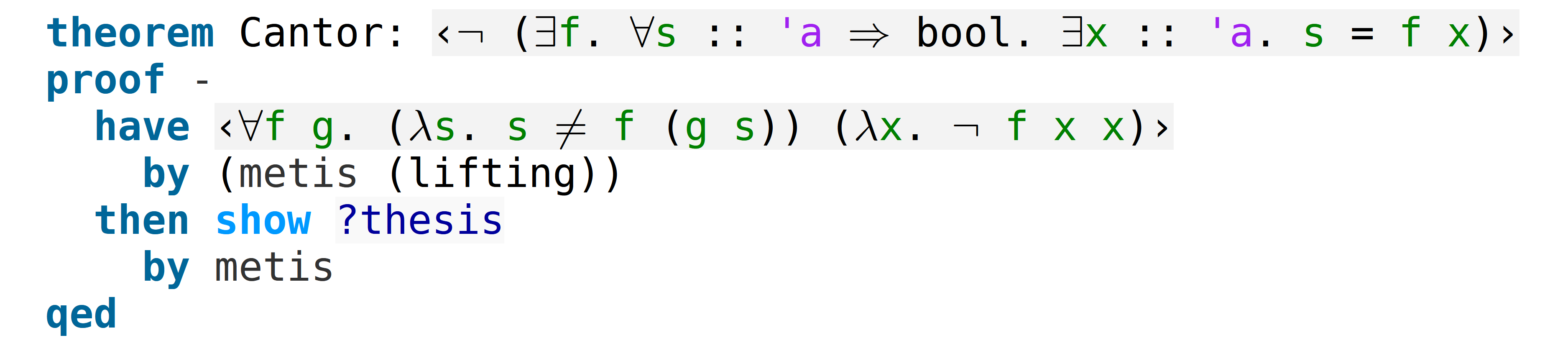}
\end{center}

In the first step, we prove that for any function $f$ from elements to sets and $g$ from sets to elements, and set $s$ that contains any $x$ such that $x$ is not in $f\ x$ (i.e. $s = ?D$), it must be the case that $s \neq f\ (g\ s)$. The reason for this is that $s$ and $f\ (g\ s)$ must disagree on the element $g\ s$. By the definition of $s$, $g\ s$ must be in $s$ if and only if it is not in $f\ (g\ s)$. This whole line of reasoning is found automatically by ``metis'' (the attribute ``lifting'' is needed).

In the second step, this is used to show the theorem. We need to show that the existence of a surjective function from a set to its power set implies a contradiction. If we assume that such a function exists, then there must also exist a function $g$ which takes a set $s$ and returns an element $x$ such that $f\ x = s$. By the property shown in the first step we know that there is an $s$ such that $s \neq f\ (g\ s)$. This, however, contradicts our definition of $g$. This line of reasoning was also found automatically by ``metis'' (the attribute ``lifting'' is not needed).

We note that the second step involves the use of the Axiom of Choice. From the fact that there for all $s$ exists an $x$ such that $f\ x = s$ we have to obtain a $g$ that picks such an element when given an $s$. This proof is therefore not possible to perform at the stage in the development where we prove Cantor's theorem above; the Axiom of Choice is introduced later on. 

\section{First-Order Logic --- Soundness and Completeness}

Our approach to teaching higher-order logic involves not only the formalization HOL\_Pure, but also implicational axiomatics in the formalization FOL\_Implicational\_Axiomatics (see Subsection 5.1) and natural deduction in the formalization FOL\_Natural\_Deduction (see Subsection 5.2), both with formal proofs of the soundness and completeness theorems in Isabelle/HOL.
Finally, in Subsection 5.3, we compare implicational axiomatics and natural deduction.

\subsection{Implicational Axiomatics}

\begin{figure}
    \centering
    \includegraphics[width=.95\textwidth]{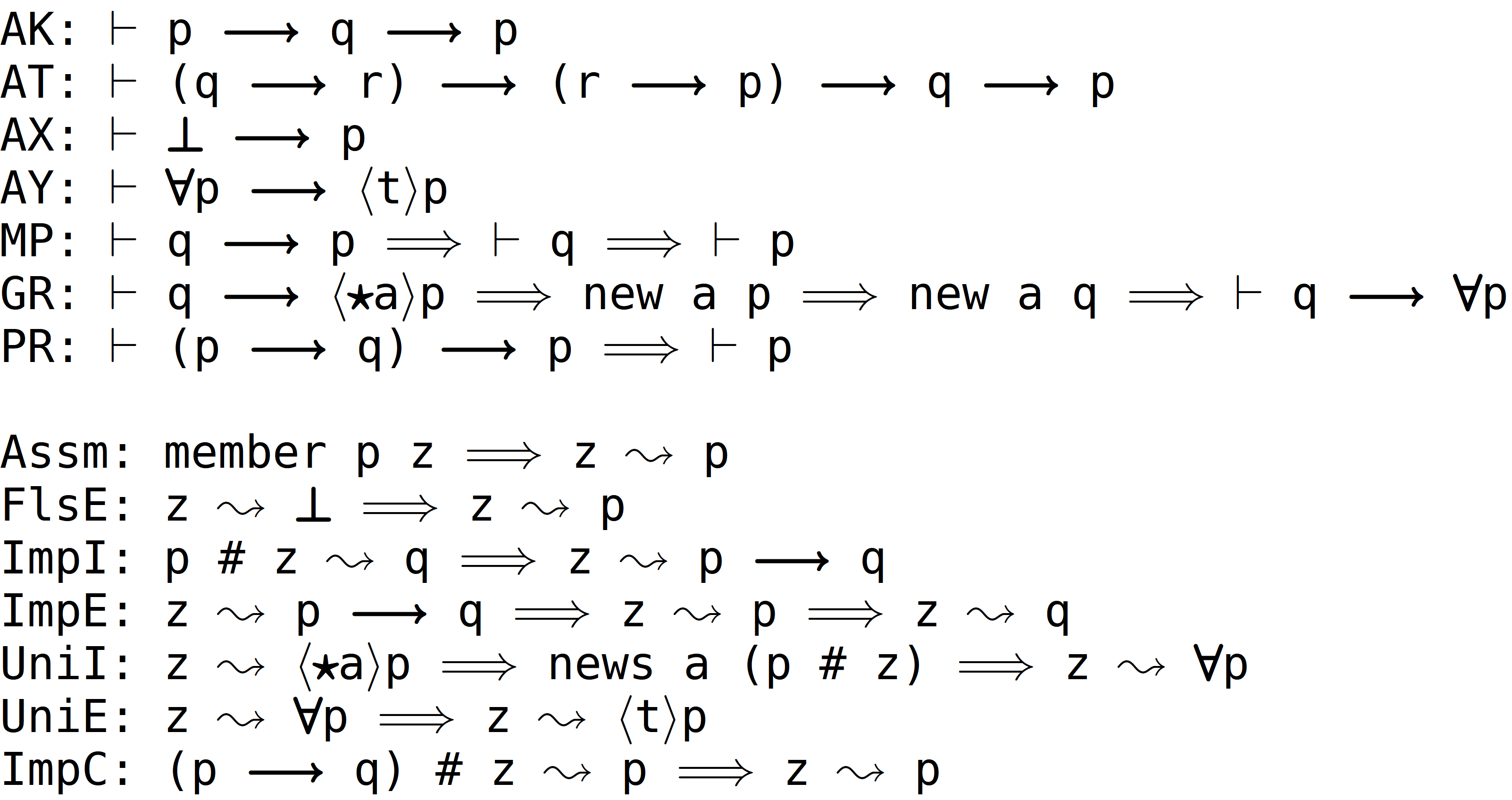}
    \medskip
    \caption{The formalization of implicational axiomatics for first-order logic (the first 7 axioms and rules) and the formalization of natural deduction for first-order logic (the last 7 rules).}
    \label{fig:fol}
\end{figure}

Natural deduction is only one of multiple styles of proof systems. Most (with some notable exceptions like resolution) can be categorized as either Hilbert style or Gentzen style, where the difference, in broad strokes, is that Hilbert-style proof systems usually have few rules and many axioms while Gentzen-style systems have few axioms and many rules. Natural deduction is a subcategory of Gentzen systems. In the natural deduction system of the next section we only need a single axiom, namely Assm. In this section we present an alternative Hilbert style axiomatic proof system for first-order logic. The proof system is shown in the top of Figure~\ref{fig:fol}. Both this and the system of the next section have been shown sound and complete in Isabelle/HOL. 

In the axiomatic system there are four axioms and three rules. The modus ponens (MP) and generalization (GR) rules are quite standard, but the Peirce's Law (PR) rule is more commonly given as an axiom. It is notable that we can still show completeness of the system where this is given as a rule instead of an axiom. If we have the axiom $((p \longrightarrow q) \longrightarrow p) \longrightarrow p$ then we can show that $((p \longrightarrow q) \longrightarrow p) \Longrightarrow p$ is admissible by a single application of modus ponens. Showing that $((p \longrightarrow q) \longrightarrow p) \longrightarrow p$ is admissible from $((p \longrightarrow q) \longrightarrow p) \Longrightarrow p$, on the other hand, requires a lengthy derivation. 

Despite the difference of approach, there is a correspondence between the individual axioms and rules of the axiomatic system and the natural deduction system of Figure~\ref{fig:fol}. We will examine these in the following. 

The AK axiom states that $p \longrightarrow q \longrightarrow p$ is valid for all $p$ and $q$. This means that if we assume $p$ (along with some passive $q$) then we can conclude $p$. This is similar to what the Assm axiom in the natural deduction system states.

The AT axiom states that implication is transitive: $(q \longrightarrow r) \longrightarrow (r \longrightarrow p) \longrightarrow q \longrightarrow p$. Such a rule is not necessary in natural deduction. The system would also be complete if we replaced this axiom with $(r \longrightarrow q \longrightarrow p) \longrightarrow (r \longrightarrow q) \longrightarrow r \longrightarrow p$ (AS), which works similarly to implication elimination under the context of a passive formula $r$. We have chosen to use AT instead of AS as we find it more intuitive and easier to explain to the students. 

The MP rule represents modus ponens: if $p \longrightarrow q$ and $p$ then $q$. In MP, modus ponens is given as a rule, which allows us to derive the validity of one formula from the validity of two other. This rule often forms the basis of reasoning within an axiomatic system. Modus ponens corresponds to implication elimination (ImpE) in natural deduction. A rule or axiom corresponding to implication introduction (ImpI) is not necessary in the axiomatic system, as all the implications necessary for building any valid formula can be introduced by the axioms. 

The AX axiom allows us to conclude anything from falsity, like FlsE in natural deduction.

The AY axiom allows us to obtain $p\ c$ for a specific $c$ from $\forall x.\ p\ x$, like UniE in natural deduction.
The generalization rule (GR) allows us to conclude $\forall x.\ p\ x$ from the validity of $p\ c$ when $c$ is fresh, similarly to UniI in natural deduction.

By adding the PR rule we extend to classical reasoning, like we do with ImpC in natural deduction.

\subsection{Natural Deduction}

In the following we describe the natural deduction system shown in the bottom-half of Figure~\ref{fig:fol}. We also describe how the rules compare to the higher-order logic system given in Section 6 and Section 7. The judgements of the proof calculus are of the form $z \leadsto p$ where $z$ contains a list of assumptions and $p$ is the formula we try to prove. 

The assumption rule (Assm) states that we can prove any formula we have assumed. This rule has no parallel in our higher-order logic, as it is implicit in the Isabelle framework; we can show $p \Longrightarrow p$ independently of our axioms.

The falsity elimination rule (FlsE) allows us to conclude anything if we have shown falsity (under a given set of assumptions). In higher-order logic it is possible to define falsity as $\forall p.\ p$, from which this property follows by the rules of the universal quantifier. In Lemma Falsity\_E we derive the rule directly corresponding to FlsE using this technique.

The implication introduction rule (ImpI) states that if one has shown $q$ while assuming $p$, then one has shown $p \longrightarrow q$.

The implication elimination rule (ImpE) states that if one has $p \longrightarrow q$ and $p$, then one has shown $q$.

The universal elimination rule (UniE) state that if one has $\forall x.\ p\ x$, then one can conclude $p\ c$ for any specific $c$.
The three rules (ImpI, ImpE and UniE) are also given as axioms (that is, they are not derived) in the higher-order logic theory. 

The universal introduction rule (UniI) here and the one for higher-order logic are defined slightly differently. Intuitively, these rules should represent that if one has shown $p\ c$ for some arbitrary $c$ (i.e. a variable which one assumes nothing about) then one has shown $\forall x.\ p\ x$. This rule is sound because whatever reason you have for concluding $p\ c$ would work for any other terms as well. When defining the higher-order logic version of this rule we can rely on the Isabelle $\bigwedge$-binder and just define the rule as $(\bigwedge x.\ p\ x) \Longrightarrow \forall x.\ p\ x$. Isabelle will allow us to continue working on the object logic inside $\bigwedge x.\ p\ x$, while guaranteeing the freshness of $x$ by binding it. In the deep-embedding approach used for first-order logic we have to do more work ourselves. First we define an Isabelle predicate for checking that a constant does not occur in a list of formulas. We can then define UniI as follows: if we can show $z \leadsto p$ after replacing $x$ with a constant $c$ in $p$ and $c$ does not occur in $z$ or $p$, then we can conclude $z \leadsto \forall x.\ p$. In the Isabelle code this looks slightly different because we use de Bruijn indices. 

The ImpC rule states that if we from $p \longrightarrow q$ can conclude $p$, then $p$ must hold on its own. This rule is not intuitionistically valid and therefore has no parallel in the higher-order logic of Section 6. After extending the logic with the classical axiom of choice and the boolean and functional axioms of extensionality, which sometimes are and sometimes are not considered a part of intuitionistic logic, in Section 7 we are able to derive it. 

\subsection{Comparison of Implicational Axiomatics and Natural Deduction}

As seen in the previous subsections, there is a substantial similarity between axiomatic first-order logic, first-order natural deduction, and higher-order natural deduction, at least when represented by the systems we have chosen.

By teaching all these systems to our students we aim to give them a broader understanding of logic and proofs and familiarize them with concepts they may encounter later. Vitally, by picking such similar systems we can really highlight the fundamental differences there are between these logics. For example that the real power of higher-order logic comes from what one allows the quantifiers to bind. Even without adding new rules this allows one to represent and prove properties that one could not in first-order logic. 

One substantial difference between the systems we define is how they treat assumptions/context. In the higher-order natural deduction system we can use Isabelle's proof state environment to save and access context without having to mention it in the rules. In the first-order natural deduction system we must work with an explicit list of assumptions. When we apply implication introduction we can save the antecedent of the implication to the assumptions and when we apply universal introduction we must check that the arbitrary variable is actually fresh with regards to the context, for example. In the axiomatic system we have no context as such, but the passive formula in some of the rules allow us to save a kind of context when using the system in practice.  

There is a further difference in what the implications in the two systems represent. In the implicational axiomatics formalization, implication is the fundamental reasoning tool used in derivations. This is why we use the same symbol, $p$, as the conclusion in all the axioms and rules. They represent different ways of deriving $p$, either by the assumptions in the same formula or by sound rules of reasoning. This focus on implication is also why we have called it implicational axiomatics. If one removed the axioms and rules containing other operators than implication (FlsE, UniI and UniE) one obtains a sound and complete proof system for classical implicational logic (i.e.\ the logic of formulas built only from implication).

In the natural deduction formalization implication is merely a logical operator to build formulas from. In this system the important logical relation is $\leadsto$, which separates the proof context (the assumptions) from the formula we are currently working to prove. The rules regarding implication in this system are either for obtaining a formula with implication (ImpI) or for using a formula with implication to show a subformula (ImpE).

\section{Higher-Order Logic --- Intuitionistic Logic}

The main difference between the natural deduction system for first-order logic and our axiomatization of higher-order logic is which terms one allows quantification over. In first-order logic we only allow quantification over the fundamental domain elements, i.e.\ those given as arguments to functions and predicates. In higher-order logic, on the other hand, we allow quantification over all terms, including predicates and functions. This increases the expressiveness compared to the first-order logic substantially, even when the same axioms are used in both systems. As an example, consider again Cantor's Theorem, which one cannot state in first-order logic. 

In the following we explain our formalization. We mostly show the definitions of our axioms and rules, and leave out the proofs of derived rules.

\begin{center}
\includegraphics[width=\textwidth]{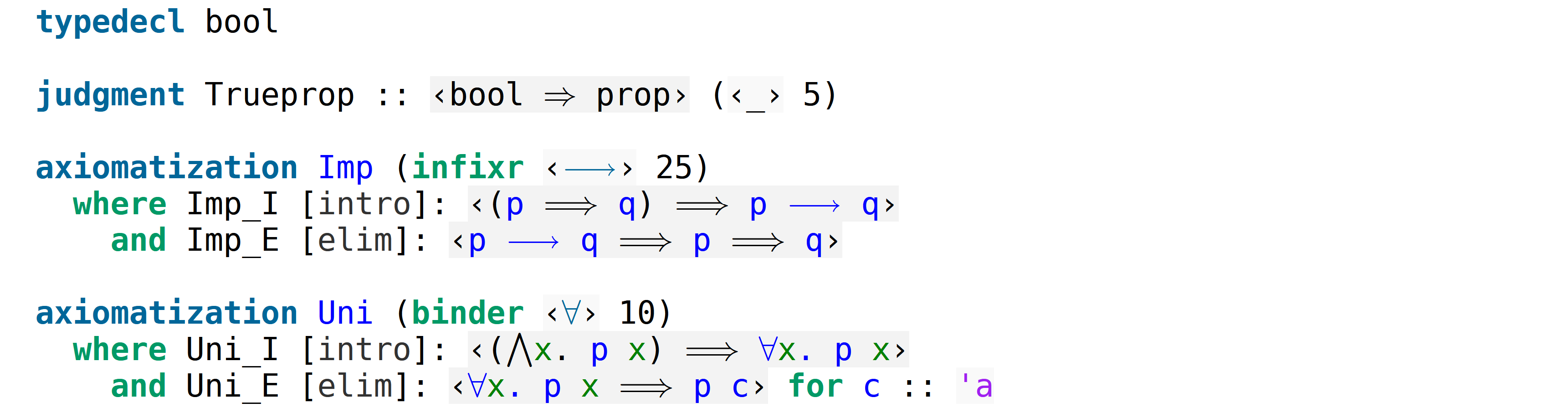}
\end{center}

The formalization starts by introducing the boolean type, which becomes the type of formulas (or facts). We then give axiomatic definitions of the object-level implication ($\longrightarrow$) and universal quantifier ($\forall$). These are defined with regards to the meta-level implication ($\Longrightarrow$) and universal quantifier ($\bigwedge$), which are defined in Isabelle/Pure.
Note that in higher-order logic, these four rules are all we need for intuitionistic logic; the other common operators can be derived as follows.

\begin{center}
\includegraphics[width=\textwidth]{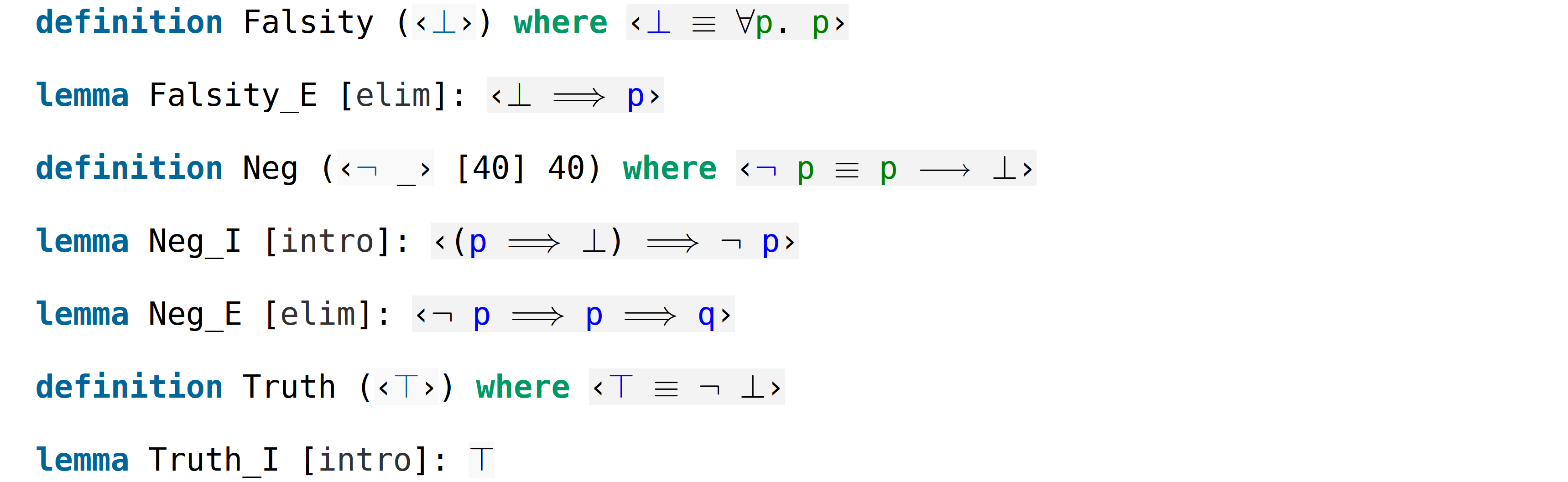}
\end{center}

We define falsity, negation, and truth. Falsity ($\bot$) is defined as $\forall p.\ p$, or ``everything is true'', which has obvious parallels to the principle of explosion. The negation of a formula $p$ is defined as $p \longrightarrow \bot$, and truth ($\top$) is defined as the negation of falsity.

We also give the introduction and elimination rules for these operators. Both the falsity elimination rule and the negation elimination rule correspond to ``from a contradiction, anything follows''. The introduction rules are as one would expect.

\begin{center}
\includegraphics[width=\textwidth]{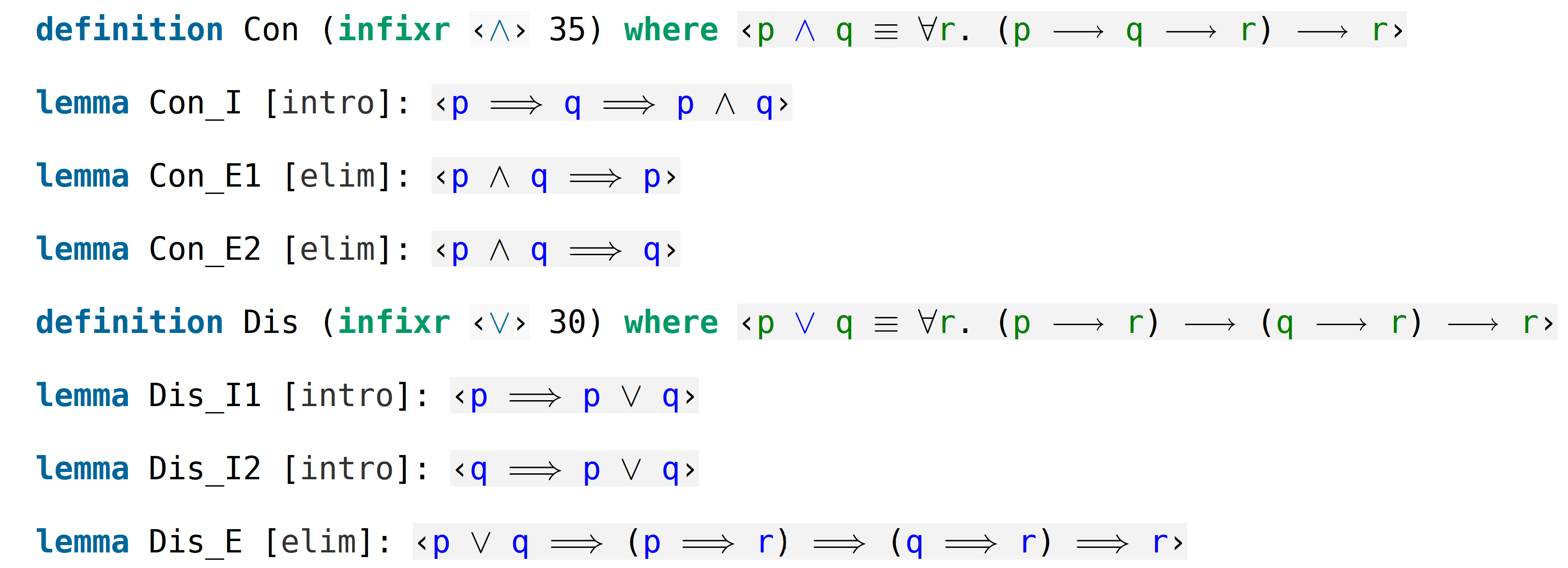}
\end{center}

Two additional propositional operators, conjunction and disjunction, are defined by the universal quantifier and implication. The alternative definitions $p \land q \equiv \neg (p \longrightarrow \neg q)$ and $p \lor q \equiv \neg p \longrightarrow q$ would be simpler, but they don't work in an intuitionistic context. That is, without extending to classical logic it is impossible to derive the introduction and elimination rules given above if one used these definitions. 

\begin{center}
\includegraphics[width=\textwidth]{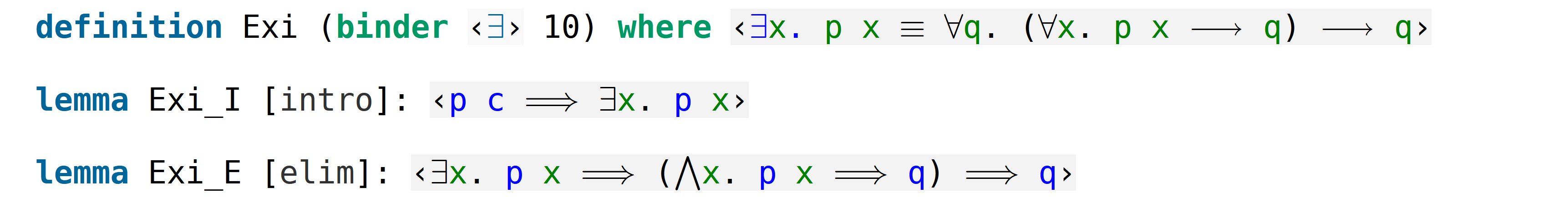}
\end{center}

The existential quantifier is also defined by the universal quantifier and implication. The elimination rule (which is very closely related to the definition) can be rewritten to the following: $(\bigwedge x.\ p\ x \Longrightarrow q) \Longrightarrow (\exists x.\ p\ x) \Longrightarrow q$ (we note that some of the parentheses are superfluous). This means that we can obtain a fixed variable fulfilling $p$ when trying to prove $q$ from the assumption that such a variable exists.

\begin{center}
\includegraphics[width=\textwidth]{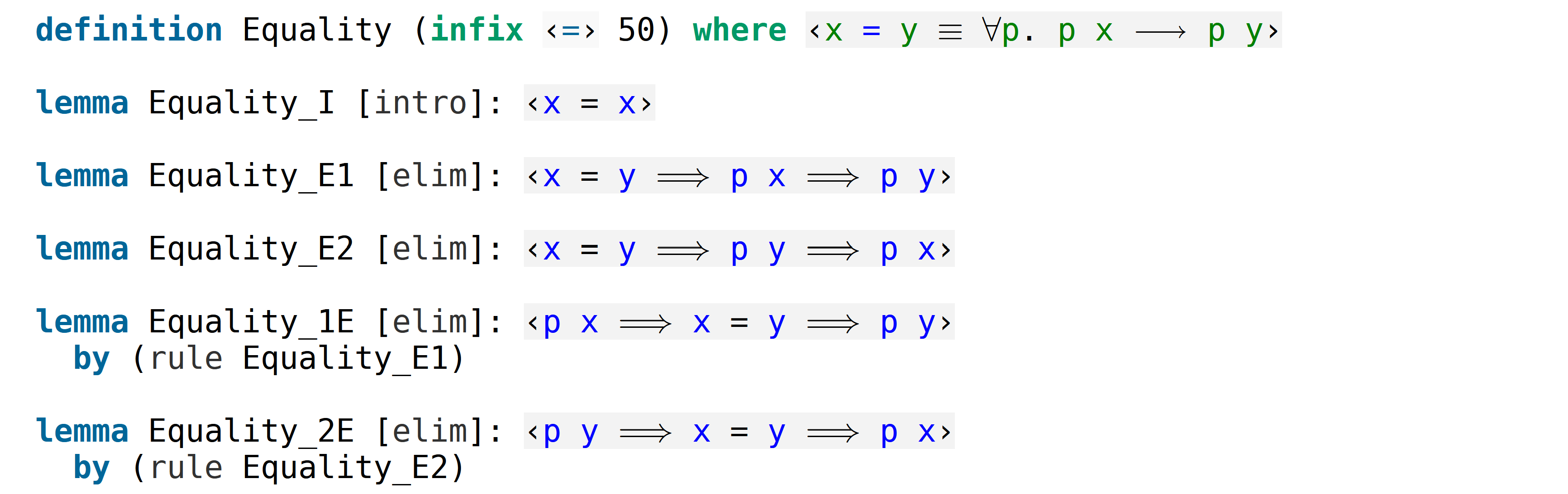}
\end{center}

We define two elements as equal if any property which holds for the first must also hold for the second. This gives rise to various substitution rules. 

These axioms, definitions, and rules are enough to prove Cantor's Theorem.

\section{Higher-Order Logic --- Classical Logic}

We will now extend the logic from the intuitionistic realm to the classical. This requires new axioms.

\begin{center}
\includegraphics[width=\textwidth]{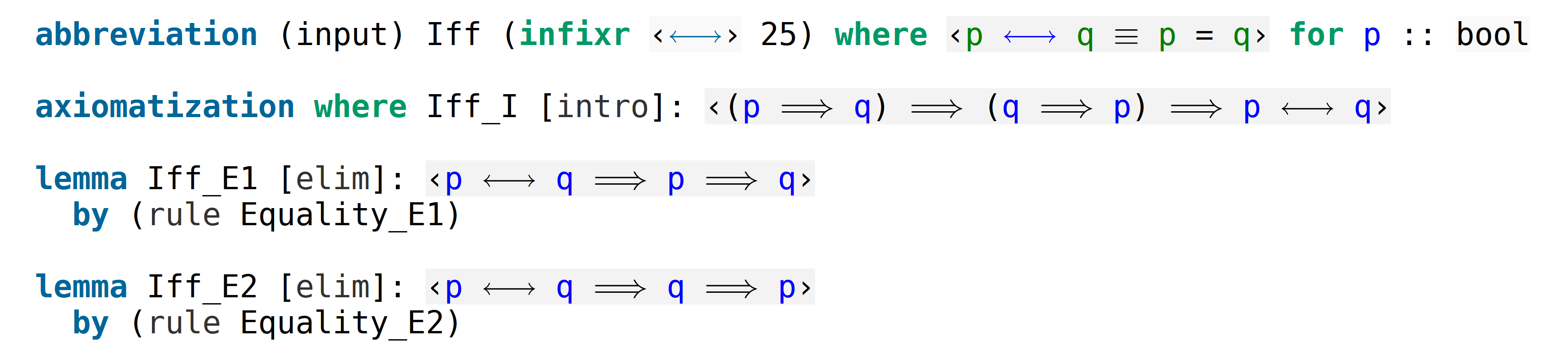}
\end{center}

The bi-implication introduction rule (Iff\_I) above and the axiom of extensionality (Extension) below are sometimes considered part of intuitionistic logic. They are, however, not derivable from the four axioms we have defined so far, and must therefore also be given as axioms.

\begin{center}
\includegraphics[width=\textwidth]{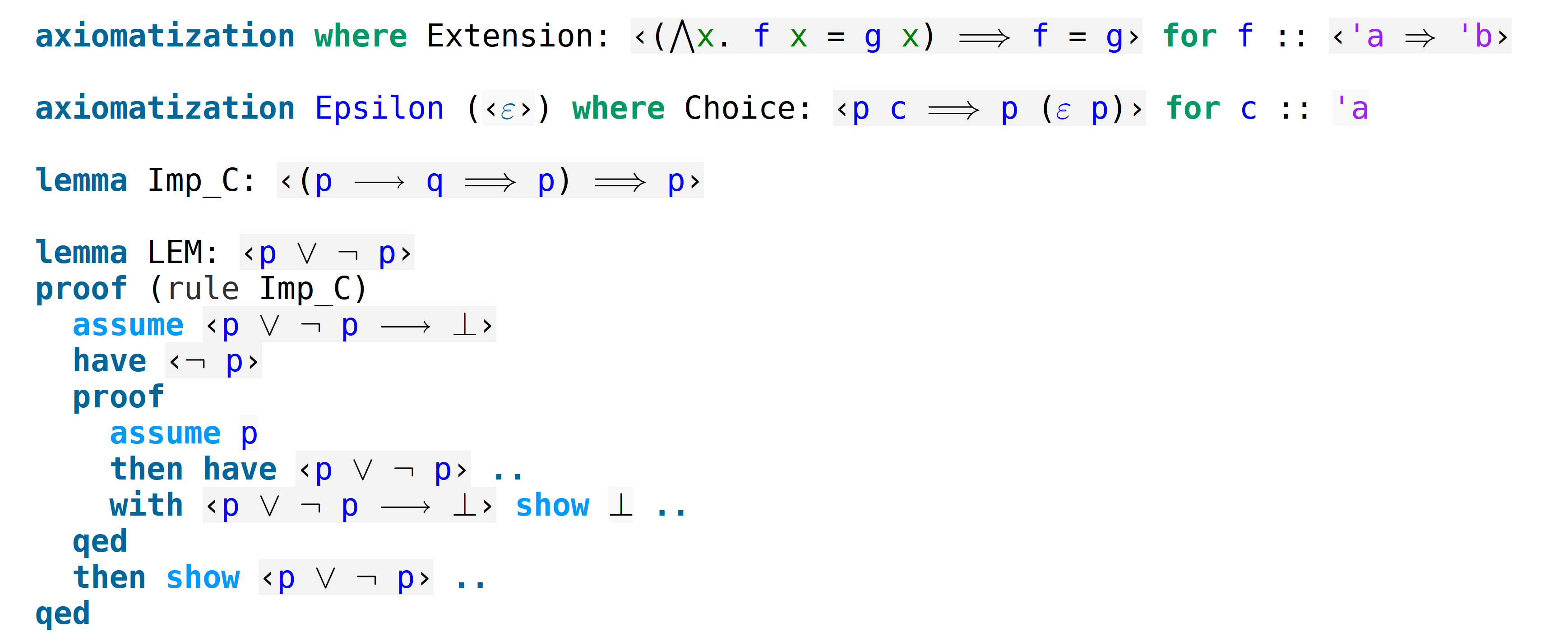}
\end{center}

Above we axiomatize Hilbert's epsilon-operator, corresponding to the axiom of choice, which finally bridges the gap to classical logic. It is now possible to prove a version of Peirce's Law (Imp\_C, where C stands for classical) which is here given as a natural deduction rule rather than a logical formula. If one instantiates $q$ with $\bot$, then Imp\_C corresponds almost exactly to the Isabelle/HOL rule named classical: $(\neg p \Longrightarrow p) \Longrightarrow p$. The proof of Imp\_C is very long, and is therefore not included. We instead show how one can use it to derive the Law of Excluded Middle (LEM).
With LEM we have classical logic as in our previous paper \cite{ThEdu21}, but now for higher-order logic and not just propositional logic.

In order to make the classical higher-order logic easier to use we can introduce the following rules known from Isabelle/HOL.

\begin{center}
\includegraphics[width=\textwidth]{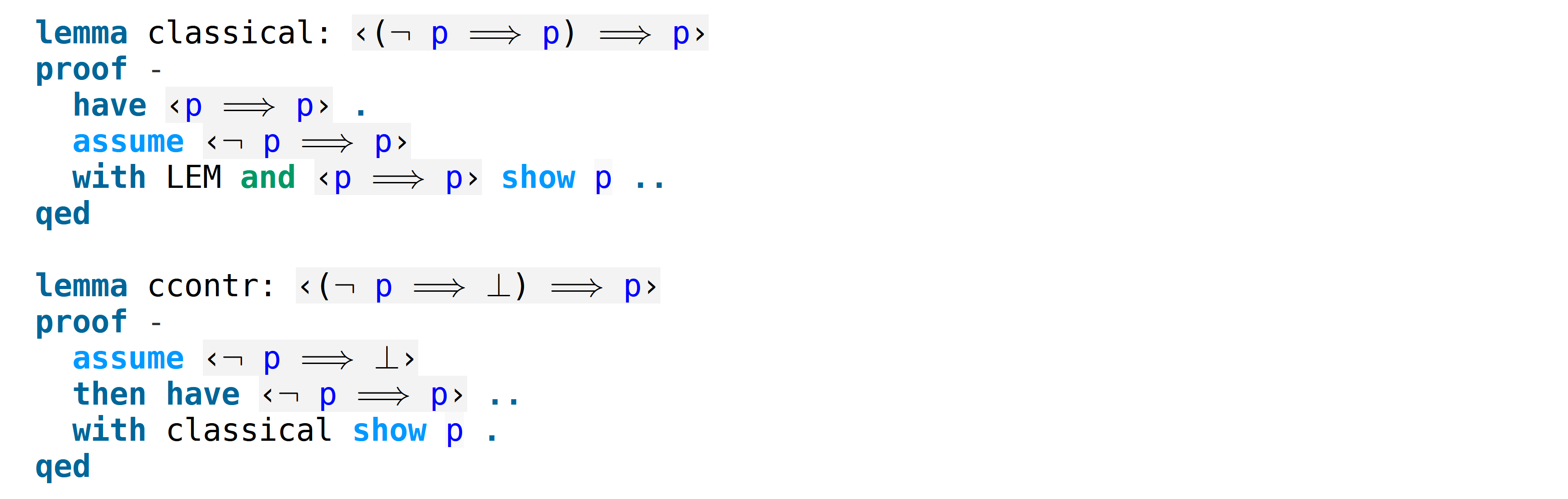}
\end{center}

Both rules are useful in the final assignment in the course where the students must prove the following formula, among others.
\begin{quote}
\emph{If every person that is not rich has a rich father, then some rich person must have a rich grandfather.}
\end{quote}
Formalization with $r$ (\emph{rich}) and $f$ (\emph{father}):
\begin{center}
$(\forall x.\ \neg\ r\ x \longrightarrow r\ (f\ x)) \longrightarrow (\exists x.\ r\ x\ \land\ r\ (f\ (f\ x)))$
\end{center}
We prefer to keep the proof secret for students.

The rule classical is also useful for students of Isabelle/HOL in proofs of, for example, the well-known Clavius's Law and Peirce's Law \cite{ThEdu21}.

The rule ccontr (for classical contradiction) is also useful in the following proof: rule (ccontr):

\begin{center}
\includegraphics[width=\textwidth]{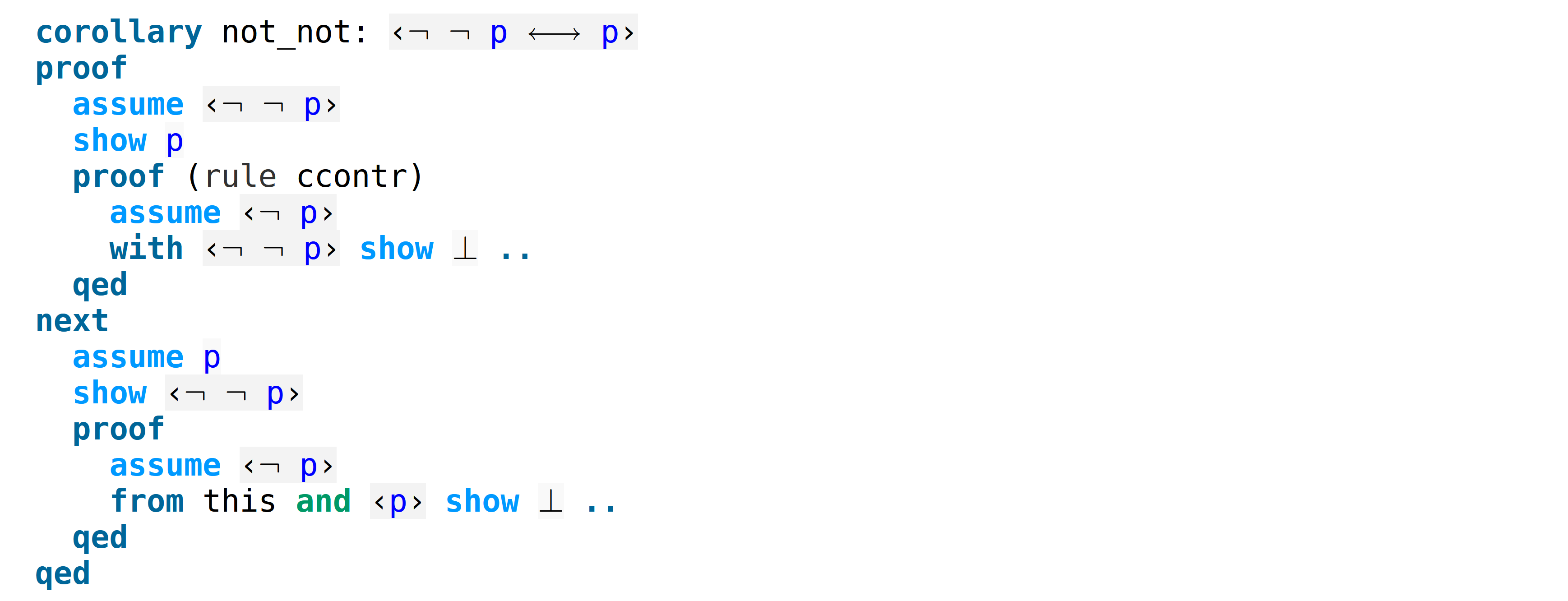}
\end{center}

We would like to emphasize that the rule ccontr (for classical contradiction) is explained in the Programming and Proving in Isabelle/HOL tutorial \cite{prog-prove}, on page 45:

\begin{center}
\medskip
\framebox{\raisebox{0pt}[27ex][3ex]{~~~~~~~~\includegraphics[width=.8\textwidth]{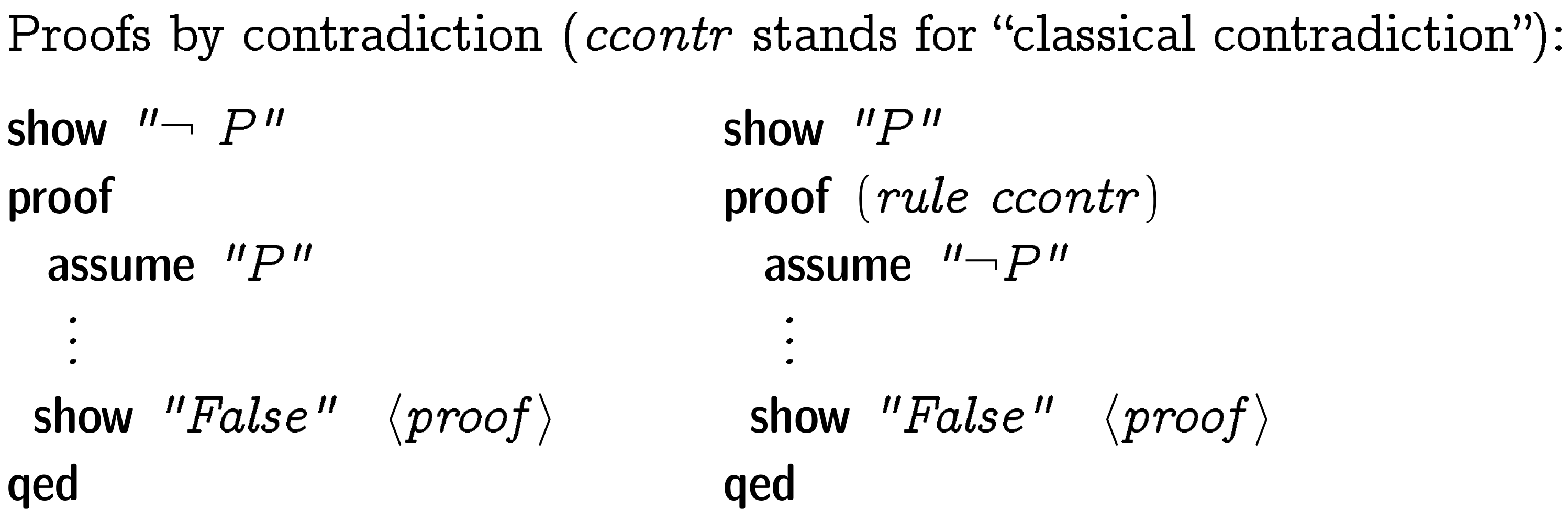}~~~~~~~~}}
\medskip
\end{center}

Note that in the tutorial the negation introduction is also called a proof by contradiction.
This might be rather confusing for students with a background in intuitionistic logic, but for most Isabelle novices it is actually a good way to view it.

\section{Further Developments and Concluding Remarks}

We now have most of the features in Isabelle/HOL but three features are missing.

The first is a simple axiomatization of the polymorphic constant for undefinedness.

\begin{center}
\includegraphics[width=\textwidth]{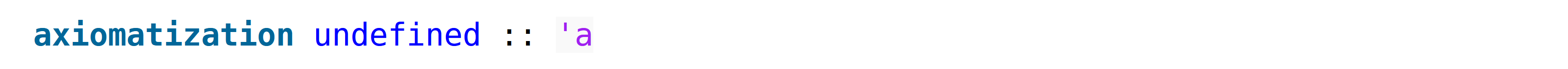}
\end{center}

Now the axiom of infinity given in two parts (Infinity\_Base and Infinity\_Step).

\begin{center}
\includegraphics[width=\textwidth]{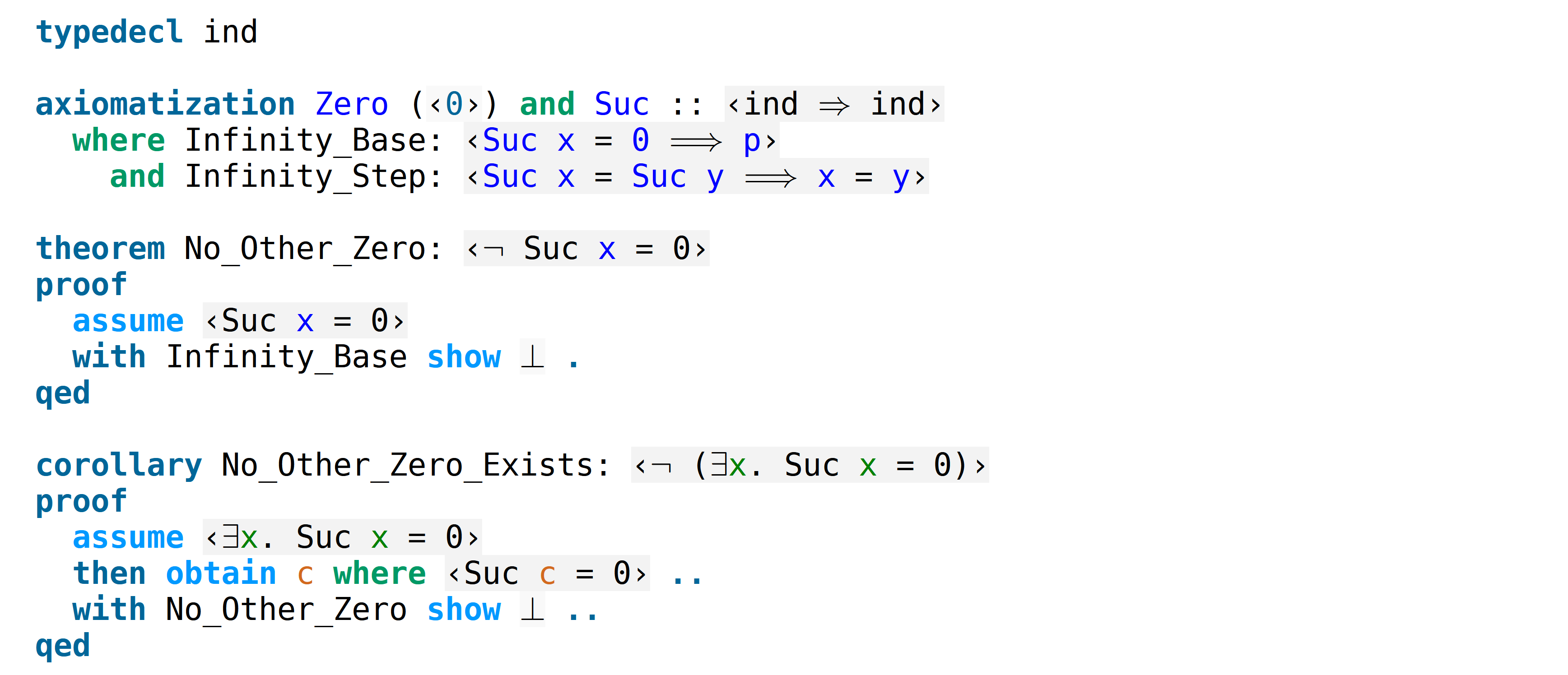}
\end{center}

Students can compare this to the similar axiomatization in the Isabelle/HOL theory Nat for natural numbers.
The later introduction of the real numbers and also the complex numbers does not require additional axioms \cite{FARMER2008267}.

Finally we introduce facilities for sets, in particular the set membership predicate ($\in$) and the axiom of comprehension (Comprehension).

\begin{center}
\includegraphics[width=\textwidth]{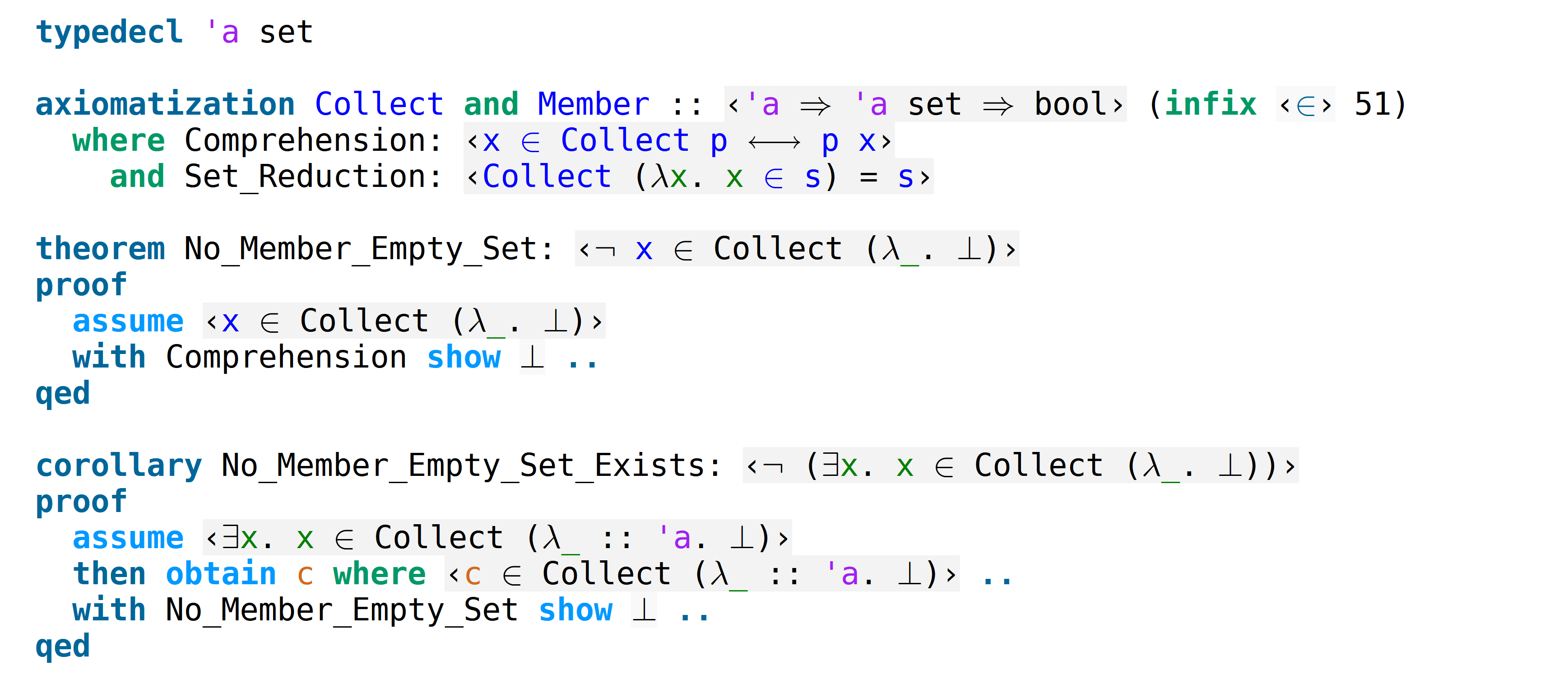}
\end{center}

Except for some small differences in names, like \emph{False} instead of $\bot$, the above theorem and corollary have identical proofs in Isabelle/HOL as documented in the following stand-alone theory.

\begin{center}
\includegraphics[width=\textwidth]{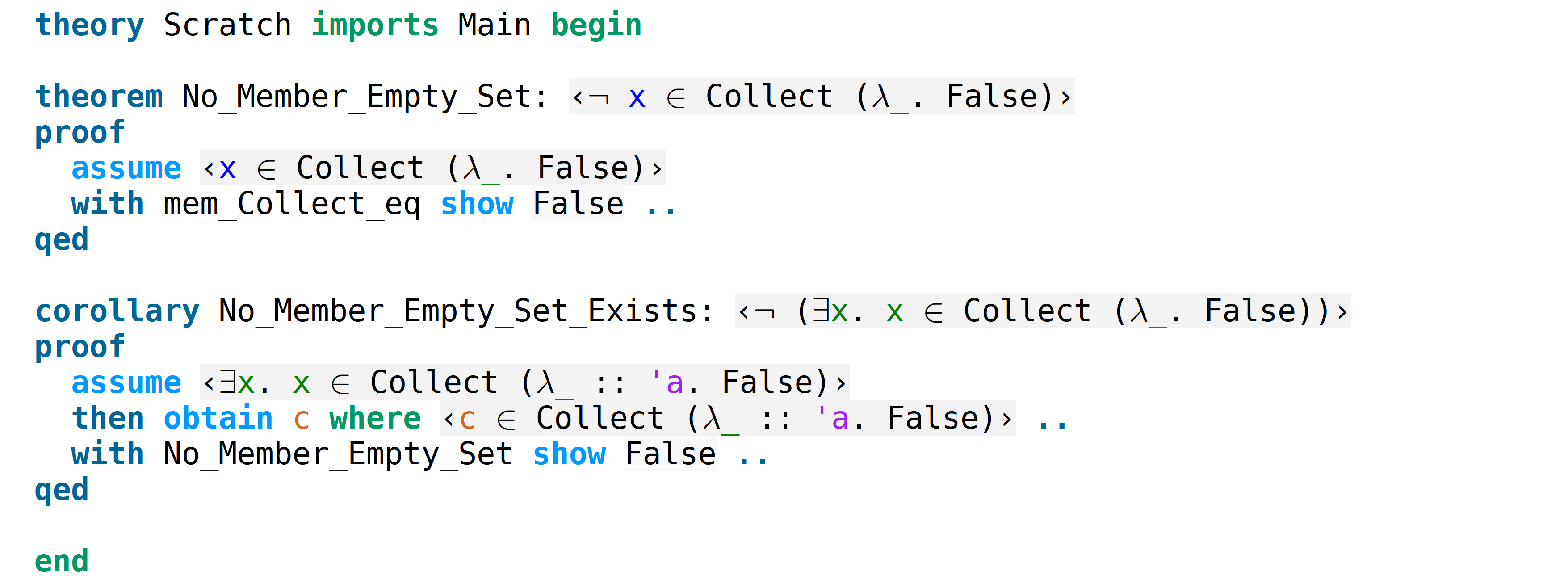}
\end{center}

The \emph{Main} theory is the main theory of Isabelle/HOL.
We note that the axiom of comprehension is called mem\_Collect\_eq in Isabelle/HOL.

\

We have found our formalization to be of great help when teaching both proving in Isabelle and in propositional logic, first-order logic and higher-order logic as such.
When stuck on a proof students can quickly go through the entire repertoire of rules to find one that might help them.
Future work includes elaboration on the overall approach, including proper teaching materials with even more explanations and examples.

As stated in the introduction, our approach has been used in a computer science course on automated reasoning in the period 2020-2023, and currently 84 students have registered in the spring 2024 course.
We plan to discuss our teaching experiences thoroughly in a forthcoming paper.
As always the Technical University of Denmark makes statistics about the course evaluations and student grades publicly available as soon as the course ends.

\

\paragraph{Acknowledgements}
Thanks to Marco Dessalvi, Frederik Krogsdal Jacobsen and Roberto Pettinau for help with the paper and the formalizations.

\

\bibliographystyle{eptcs}
\bibliography{references}

\end{document}